# Effect of epitaxial strain on the magneto-electric coupling of YMnO$_3$ thin films

A. K. Singh and S. Patnaik
School of Physical Sciences
Jawaharlal Nehru University New Delhi 110067 India

M. Snure and A. Tiwari
Nanostructured Materials Research Laboratory
Department of Materials Science and Engineering
University of Utah, Salt Lake City, UT 84112 USA

We report synthesis of phase pure multiferroic YMnO$_3$ thin films on sapphire (0001) with conducting ZnGaO buffer contact layer.  Films were prepared by using pulsed laser deposition technique and characterized using x-ray diffraction (XRD), Scanning electron microscopy (SEM), Energy dispersive absorption spectroscopy (EDAX) and magnetic field dependent dielectric measurement techniques.  Structural characterizations indicated phase purity and epitaxial nature of the films.  The dielectric response indicated an anomaly in dielectric constant ε and tanδ in the vicinity of 30 K, well below the bulk Néel temperature ~ 70 K.  This anomaly in ε (T) and tanδ  and its magnetic field dependence is explained as an influence of strain due to lattice mismatch between the substrate and YMnO$_3$ film.  A substantial enhancement in magnetocapacitance was also observed for magnetic field applied parallel to *ab* plane of the film.  Our results show that it is possible to tune the multiferroic property of YMnO$_3$ via changes in ferroelastic route.

# I. INTRODUCTION

Multiferroics are rare materials that possess two or more switchable states such as polarization, magnetization, and/or strain and exhibit interdependence between the corresponding ordered phases.[1-8] In the recent past there has been a renaissance in research on multiferroic systems primarily due to two reasons; the discovery of improper ferroelctricity in spin frustrated systems[9-11] and the observation of strain induced enhancement in ferroelectric Curie temperature and polarization in thin films of complex oxides.[12-15] Both these developments enable potential applications of multiferroic materials in information storage and spintronics such as magneto-electric sensors, magneto-capacitive devices, and electrically driven magnetic data storage. Particular emphasis has been placed on hexagonal $YMnO_3$ where a microscopic origin of multiferrocity is clearly enunciated.[16-18] Bulk $YMnO_3$ shows a ferroelectric transition $T_E \sim 900$ K and antiferromagnetic ordering $T_N \sim 70$ K and the dielectric constant ($\varepsilon$) and loss factor ($\tan\delta$), that reflect the electric order, show magnetic field dependence across the antiferromagnetic transition and thus establish multiferrocity below 70 K. An open question that is of current interest is whether this onset temperature can be increased or decreased by introducing strain into the system. In this paper we establish that strain can indeed play a significant role in determining the magnetic and electric ordering temperature in lattice mismatched $YMnO_3$ thin films. We have also investigated how magneto-electric coupling is affected due to tensile strain. Our results indicate that it is possible to tune the magnetic and electric ordering parameters in multiferroic materials by the means of strain.

A major series that show ferroelectricity along with magnetic ordering consist of materials of the general formula $RMnO_3$ (R = Y, Lu, Ho, Er, Tm, Sc, Yb etc.). From the structural point of view, magnetic ordering occurs in both hexagonal (space group $P6_3cm$) and orthorhombic (space group Pnma) manganites, whereas ferroelectric ordering occurs only in hexagonal phase which is noncentrosymmetric.[6] Hexagonal $RMnO_3$ compounds have antiferromagnetic ordering with the Néel temperature $T_N$ < 70-130 K, and ferroelectric ordering at much higher temperature $T_E$ < 600-900K.[16-18] Particularly, in $YMnO_3$ the origin of multiferrocity is explained as follows. In hexagonal $YMnO_3$ each Mn ion is surrounded by three in-plane and two apical oxygen ions. These $MnO_5$ blocks are connected two dimensionally through their corners, and the triangular lattice of $Mn^{3+}$ ions with S=2 are formed. The $Mn^{3+}$ ions interact antiferromagnetically and lead to geometrical frustration (GF) in the system. Due to this, $MnO_5$ blocks tilt leading to displacement of $Y^{3+}$ ions along c-axis and lead to consequent ferroelectric polarization.[16,17] Thin films and single crystals of $YMnO_3$ show that these materials are highly anisotropic in terms of magnetic and electrical properties.[6] The question that we are seeking to answer is whether this tilting of $MnO_5$ polyhedra can be altered controllably by lattice strain.

Various techniques, including chemical solution deposition (CSD)[19], sol-gel, and metal-organic chemical vapor deposition (MOCVD)[20], molecular beam epitaxy (MBE)[21], sputtering[22] and pulsed laser deposition (PLD)[23] etc. have been employed successfully to fabricate $YMnO_3$ thin films. However, in order to make $YMnO_3$ films with good potential application the control of crystal orientation is very important. This is because the ferroelectric polarization appears along the c-axis.[24] Moreover, from the standpoint of fabricating actual multi-functional devices, it will be needed to attach top and bottom conducting electrodes to these films. Though the fabrication of the top contact is quite straightforward, the fabrication of the bottom contact offers a great challenge. To maintain the high crystal quality of the YMnO$_3$ film, the bottom contact layer itself need to be epitaxial with the YMnO$_3$ as well as with the sapphire substrate. In this work we have found that Ga doped ZnO is a very good candidate for making bottom contact to YMnO$_3$ when deposited on sapphire (0001) substrate.

As is well known, in thin film samples grown on heteroepitaxial substrates, a significant amount of stress is always present. The origin of stress can either be extrinsic, caused by the deposition of the film at one temperature and measurement at another temperature or intrinsic, resulting from the microstructure of the film or a misfit in an epitaxial relation between the film and the substrate.[25] Recently Eerenstein *et al.* observed that magnetic, resistive, and magnetoelectric effects are primarily attributed to the strain induced coupling at the interface.[26] In this paper we have investigated the characteristics of the YMnO$_3$ film grown on sapphire/Zn$_{0.99}$Ga$_{0.01}$O possessing tensile stress. The crystallinity and the surface morphology of the YMnO$_3$ film were found to be dependent on the lattice matching between the substrate and YMnO$_3$.

## II. EXPERIMENTAL DETAILS

YMnO$_3$ and ZnGaO films were deposited on sapphire substrate by PLD technique starting from high purity single phase bulk targets prepared using a low temperature sol-gel technique. A KrF pulsed excimer laser (248nm wavelength and pulse width of 25ns) was used for ablating the targets. In the first step of the deposition, the deposition chamber was evacuated to a base pressure of 10$^{-6}$ torr and a 100 nm thick epitaxial layer of Zn$_{0.99}$Ga$_{0.01}$O was deposited on the sapphire. During the above deposition process, temperature of the substrate was kept at 600 $^0$C, energy density was 2 J/cm$^2$ and a pulse repetition rate of 10 Hz was used. In the second step of deposition, YMnO$_3$ film was deposited on Zn$_{0.99}$Ga$_{0.01}$O coated sapphire (0001) substrates with an energy density 3 J/cm$^2$ at a pulse repetition rate of 10 Hz and a substrate temperature of 700°C. Approximately 200 nm thick films were obtained after a deposition time of 10 min.

The structural properties and crystallographic orientation of YMnO$_3$ thin films were characterized using Philips X-pert x-ray diffraction (XRD) system. The capacitance (C) and the loss factor (tanδ) were measured with a QUADTECH 1920 precision LCR meter. To characterize the dielectric properties, gold electrode of nominal dimension 3×2 mm$^2$ was deposited at room temperature on top of the film using an appropriate shadow mask. It is to be noted that 1% Ga doped ZnO substrate is highly conducting (room temperature resistivity ~ 1.44 ×10$^{-4}$ Ω-cm) and that formed the second terminal for the dielectric measurement.[27] Capacitance-voltage (C-V) measurement has been done using Agilent HP 4285A Precision LCR meter at the frequency 100 kHz. The resistivity of the insulating film was measured by using a constant voltage source in a two-probe configuration. The magnetic DC susceptibility (χ) was determined by Quantum Design (SQUID) magnetometer.

### III. RESULTS AND DISCUSSION

Fig. 1(a) shows the XRD patterns of YMnO$_3$ thin film on Ga doped ZnO substrate. In the 2θ range 20 to 100 we could see only the peaks corresponding to (0006) and (00012) planes of sapphire, (0002) and (0004) planes of ZnGaO and (0002) plane of YMnO$_3$ indicating the highly c-axis aligned nature of the film. Electron dispersive absorption (EDAX) spectrum of YMnO$_3$ thin film grown on Zn$_{0.99}$Ga$_{0.01}$O/sapphire substrate is shown in fig. 1(b). Only the peaks corresponding to elements of the YMnO$_3$ film and sapphire/Zn$_{0.99}$Ga$_{0.01}$O are observed indicating the high phase-purity of the material. The ferroelectric behavior of the YMnO$_3$ films was confirmed by room temperature C-V characteristics. The butterfly nature of the C-V curves [fig.2a] suggests ferroelectric behavior at room temperature. These measurements were done by applying small alternating voltage of amplitude 10 mV at a frequency of 100 kHz and a bias voltage of 5 V. This C-V characteristic has a hysteretic nature, which indicates ferroelectric polarization switching behavior. These results show that remnant polarization of YMnO$_3$ film induces charge compensation on the ZnO substrate surface. However the memory retention property is very poor. This is probably due to the compensation of remnant polarization by a positive charge. In an attempt to characterize the antiferromagnetic (AFM) transition associated with the YMnO$_3$ film we have examined the film by SQUID measurements. It is to be pointed out that the signal from such measurement is extremely weak and generally shrouded by large paramagnetic contribution from the substrate. Nevertheless at an external field of 0.1T we observe a clear reversal in magnetic moment, characteristic of AFM ordering at T ~ 30 K (fig. 2b). One other crucial aspect need to be established for genuine magnetocapacitive effect is to rule out contribution from magnetoresistance.[28] The response to the applied electric field contains both capacitive (dielectric) and resistive (leakage) term. To check whether the observed anomaly in the dielectric constant is coming from magnetoresistance, resistivity measurements in zero field as well as in presence of 3 T magnetic field were performed. The DC resistivity ($\rho$) was found to increase exponentially as temperature decreased (fig. 2c). Even in the presence of 3T magnetic field no magneto-resistance is observed that confirms that any anomaly appearing in the dielectric constant measurement is dominantly of capacitive origin.

We next turn to the temperature and magnetic field dependence of the real and imaginary part of dielectric constant for the YMnO$_3$ film shown in fig.3. While the tan$\delta$ as a function of temperature is shown in main panel, in inset a and b, tan$\delta$ and dielectric constant are plotted in the presence of magnetic field. The dielectric constant increases with temperature and shows a sharp upturn at T ~ 30 K. In tanδ, on the other hand, there is clear evidence of an inverse S-shaped anomaly around the same temperature window. Such anomaly has been identified as the signature of the onset of antiferromagnetic transition in a multiferroic system.[29] In polycrystalline $YMnO_3$ samples, this inverse S-shape anomaly is observed near $T_N$ (~ 70 K). The decrease in the temperature for anomaly in tanδ from 70 K to ~ 25 K is clearly due to the strain produced due to lattice mismatch of ZnGaO/sapphire substrate and the $YMnO_3$ film. It is to be noted that the lattice parameters for hexagonal $YMnO_3$ and ZnO are **a** = 3.06 Å and **a** = 3.25 Å respectively that indicates a basal lattice mismatch of 5.8%. On the application of a magnetic field of 3 T, both ε and tanδ decrease. Further, this decrease is most pronounced when the field applied is parallel to ab-plane. As shown in inset a and b, for field applied perpendicular to the ab-plane, no significant effect on either ε or tanδ is observed. These measurements show the anisotropy in the magnetic field dependent dielectric property of $YMnO_3$. In fig 4 we show the frequency response of loss (tanδ) as a function of temperature for the $YMnO_3$ film. As we increase the frequency the dielectric constant decreases and the shoulder in tanδ broadens and shifts towards higher temperature. The loss data in fig. 4 reveals the occurrence of a transition over the 22 K-30 K range. We have also done measurements on polycrystalline $YMnO_3$ (shown in the inset of fig. 4) to confirm the signatures of magnetoelectric coupling in the case of randomly oriented grains. In zero field measurement a clear anomaly at T ~ 70 K is observed and after application of 4 T magnetic field, this anomaly is suppressed. We note that while the magnetocapacitance ((ε(H,T) - ε(0,T)/ ε(0, T)) ×100) for polycrystalline $YMnO_3$ pellet is ~ 0.3 at 4 T, in case of epitaxial $YMnO_3$ film this increases significantly to ~ 5.5 at 3 T. This order of magnitude enhancement in strained thin film has substantial ramification for device applications.

While a microscopic understanding of the effect of strain in controlling the onset of multiferrocity needs further study, in the following we offer a qualitative understanding of our main results. In case of normal ferroelectrics, a broad transition is assigned to the presence of grains and termed "relaxor behavior". The presence of charged oxygen vacancies, and short range polar-regions can give rise to such features leading to frequency dependence of dielectric constant.[30] One possible scenario relates to the movement of domain walls. We note that although our film is ferroelectrically ordered, it does not constitute a single domain. It has been experimentally shown that at low temperature, both ferroelectric and antiferromagnetic domains can coexist and coincide with one another.[24] The application of electric or magnetic field will induce movement of domain walls that would cost energy.[31] This is the reason for decrease in magnitude of dielectric constant ($\varepsilon$) and loss ($\tan\delta$) after the application of magnetic field. In strained ferromagnetic films, the variation of Curie temperature has been attributed to change in the bond lengths and bond angles of Mn-O octahedral and splitting of $e_g$ levels.[12-13, 32] In our case too a similar behavior can be expected leading to change in magnetic ordering temperature as indicated in the SQUID and field dependent dielectric measurements.

## IV. CONCLUSIONS

In conclusion, we report synthesis of epitaxial and highly oriented $YMnO_3$ thin films on Ga doped ZnO buffer on top of (0001) sapphire using pulsed laser deposition technique. Our results show that the magnetic and dielectric property of the film depends on the lattice strain between the substrate and $YMnO_3$ film. The inverse S- shape anomaly in loss factor, accompanying antiferromagnetic ordering, shifts to ~30 K as compared to bulk $YMnO_3$ which is at ~ 70 K. We identify anisotropic magnetocapacitance and an order of magnitude increase in its magnitude in thin films as compared to bulk YMnO$_3$. Our results elucidate that it is possible to tune multiferroic properties by lattice strain in case of hexagonal YMnO$_3$. The next challenge would be to find the right magnitude of strain so as to increase the multiferroic regime higher than liquid nitrogen temperature.

**ACKNOWLEDGEMENTS**

We thank Dr. P. K. Mishra (BARC, Mumbai) for the SQUID measurements and very useful discussions. SBP thanks the Department of Science of Technology, Government of India, for the support under the FIST program. A.K.S. thanks CSIR, India for research fellowship.

**Figure Caption**

Fig.1 (a) X-ray diffraction (XRD) Intensity - 2θ pattern of YMnO$_3$ thin film grown on Zn$_{0.99}$Ga$_{0.01}$O/sapphire substrate. (b) Energy dispersive spectrograph (EDAX) of YMnO$_3$ film.

Fig.2 (a) Room temperature capacitance – voltage (C-V) characteristics of YMnO$_3$ film at 100 kHz. The butterfly nature of C-V curves suggests a weak ferroelectric behavior at room temperature. (b) Temperature dependence of magnetization of YMnO$_3$ film is shown. The zero field cooled data is taken in the warming cycle with an external magnetic field of 0.1 T parallel to c-axis. An antiferromagnetic downturn is observed near 30 K. (c) Temperature dependence of resistivity (ρ) at $\mu_0H$ = 0 T (●) and 3 T (▲). No trace of magnetoresistence is observed.

Fig.3 The loss factor (tanδ) as a function of temperature at 123 Hz is plotted. An inverse S-shape anomaly is observed which reflects the onset of antiferromagnetic ordering in the system. Inset (a) shows tanδ and (b) shows dielectric constant (ε) as a function of temperature and magnetic field. The data are taken at 123 Hz and $\mu_0H$ = 0 T (■), 3 T ⊥ to ab-plane (●), and 3T ∥ to ab-plane (▲).

Fig.4 Frequency and temperature dependence of tanδ. Symbols a-h represent data taken at 31, 53, 73, 123, 523, 1023, 5023, and 10023 Hz respectively. At lowest frequency (31 Hz) the inverse S anomaly is clearly evident. As we increase the frequency the magnitude of anomaly in tanδ decreases and at 10023 Hz the anomaly is barely visible. Inset shows the dielectric constant a function of temperature for a polycrystalline YMnO$_3$ pellet in zero field and at 4 T magnetic field. The anomaly in dielectric constant corresponding to antiferromagnetic transition is seen at T ~ 70 K.

**Fig.1**

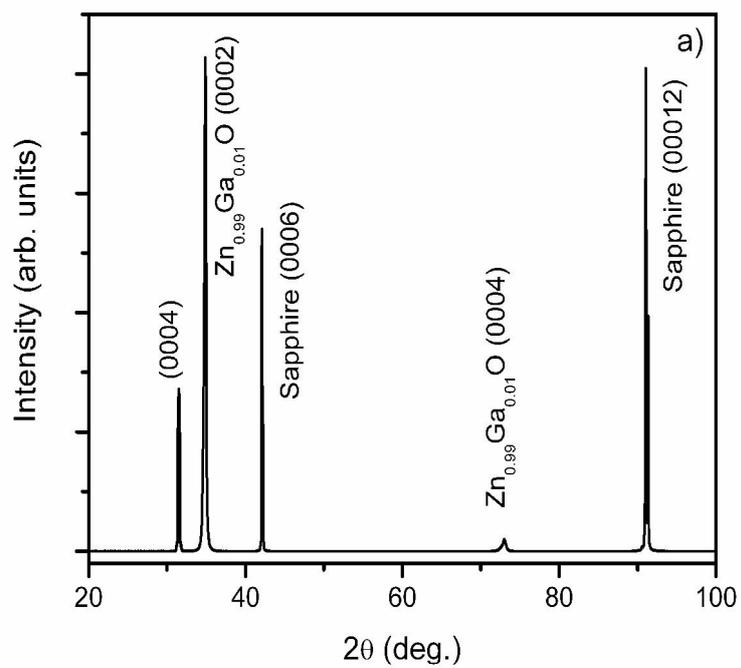

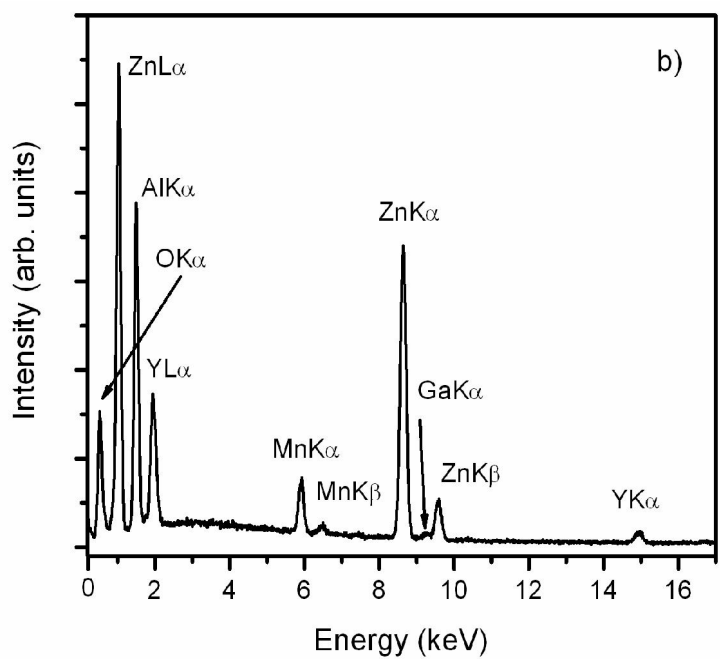

**Fig.2**

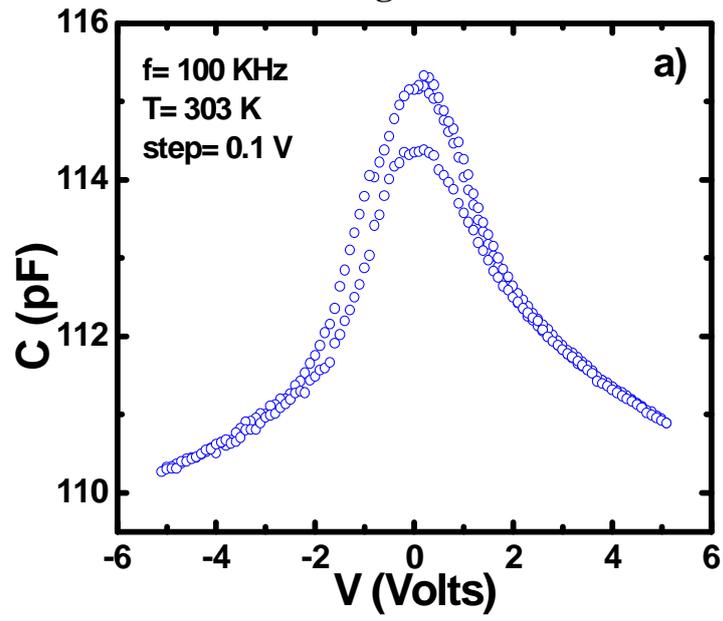

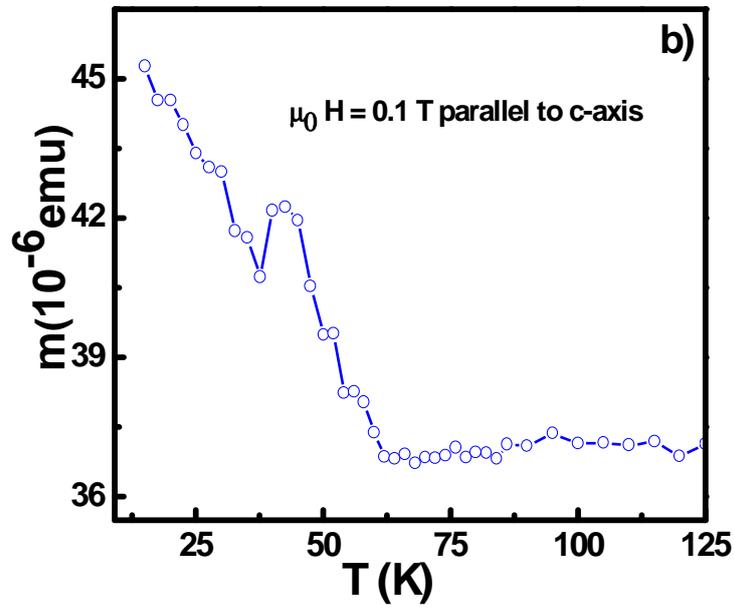

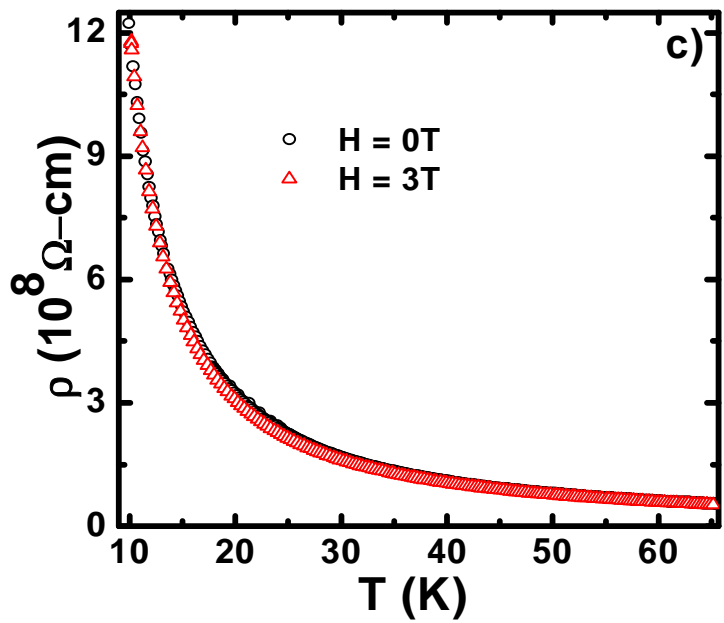

**Fig.3**

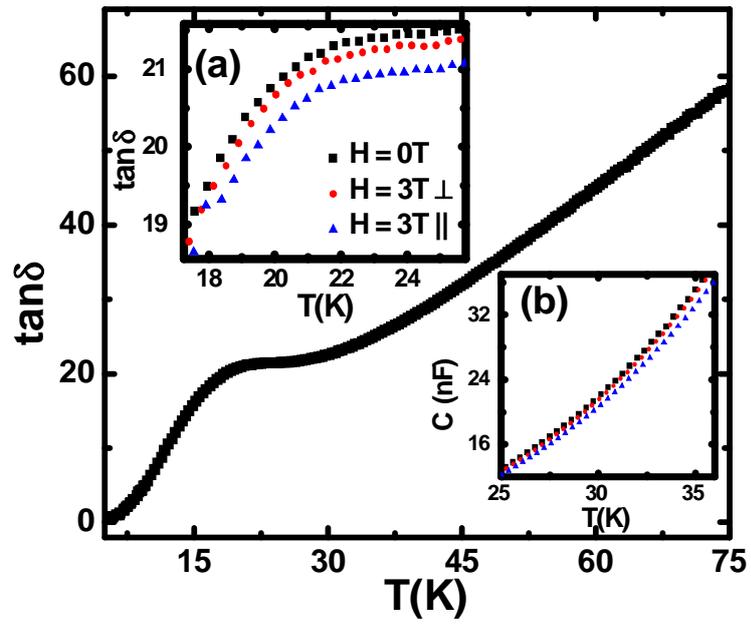

**Fig. 4**

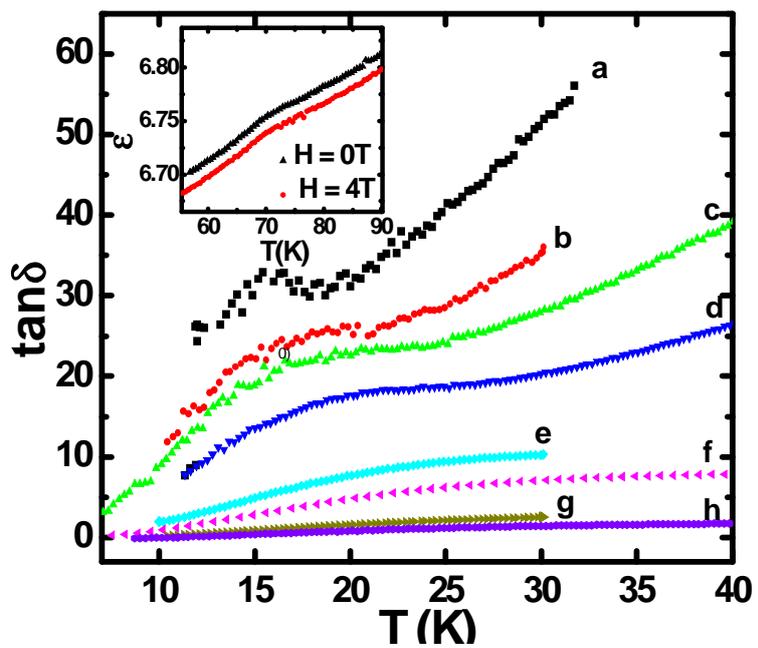